\documentclass[12pt,a4paper]{article}

\usepackage{amsmath,amssymb,psfrag,graphicx}

\begin{document}

\begin{titlepage}
\begin{center}
\hfill LMU-ASC 20/06\\
\hfill MPP-2006-27\\
\hfill {\tt hep-th/0603166}\\
\vskip 10mm

{\Large \textbf{On supersymmetric Minkowski vacua in IIB orientifolds}}
\vskip 16mm

\textbf{Daniel Krefl$^{a}$ and  Dieter L\"ust$^{a,b}$}

\vskip 4mm
$^{a}${\em Arnold Sommerfeld Center for Theoretical Physics\\
Department f\"ur Physik,
Ludwig-Maximilians-Universit\"at M\"unchen, 80333 Munich, Germany}\\

\vskip 4mm
$^{b}${\em Max-Planck-Institut f\"ur Physik, 80805 Munich, Germany}\\
\vskip 4mm

{\tt krefl,luest@theorie.physik.uni-muenchen.de}
\end{center}
\vskip .6in
\begin{center} {\bf ABSTRACT } \end{center}
\begin{quotation}\noindent
Supersymmetric Minkowski vacua in IIB orientifold 
compactifications based on orbifolds with background fluxes 
and non-perturbative superpotentials are investigated. 
Especially,
microscopic requirements and difficulties to obtain such vacua are discussed. 
We show that orbifold
models with one and two complex structure moduli and 
supersymmetric 2-form flux can be successfully stabilized to such vacua. 
By taking additional gaugino condensation on fixed 
space-time filling D3-branes into account also models without 
complex structure can be consistently stabilized to Minkowski vacua.
\end{quotation}

\vfill
\end{titlepage}

\eject

\section{Introduction}

Moduli stabilization in superstring theory has been an 
unsolved problem for a long time. However, during recent time significant 
progress has been made. An important step was to recognize the 
importance of flux backgrounds \cite{Grana:2005jc} 
for moduli stabilization issues. 
E.g. turning on 3-form fluxes in type IIB orientifolds 
generates a potential for the axion-dilaton and the complex structure moduli. 
However, in general K\"ahler moduli stay unstabilized.
To overcome this difficulty, KKLT \cite{Kachru:2003aw} 
proposed to lift the remaining flat directions by considering non-perturbative effects. 
In particular, gaugino condensation in super Yang-Mills theory of D7-branes wrapping 
internal 4-cycles or instanton effects via euclidean D3-branes also 
wrapping 4-cycles may give a proper non-perturbative term to the superpotential to lift the flat directions. 
Alternatively, one might also consider the 
possibility that $\alpha'$ and perturbative 
effects might be sufficient to lift the flat directions 
\cite{vonGersdorff:2005bf,Balasubramanian:2005zx,Berg:2005yu}.
After having stabilized all moduli to an AdS space the KKLT
scenario in addition proposes 
to uplift the AdS vacuum to a dS vacuum by 
introducing $\overline{D3}$-branes.

Despite this remarkable success the 
situation concerning full moduli stabilization is still not finished. 
More detailed investigations and applications of the KKLT scheme 
to more complicated models quickly uncovered that the 
consistency of the scheme is strongly model dependent. 
Specially, the creation of non-perturbative potentials for the 
K\"ahler moduli  strongly depends on the fluxes 
and the topology of the compactification manifold 
\cite{Witten:1996bn,Gorlich:2004qm,Kallosh:2005gs,Martucci:2005rb,Bergshoeff:2005yp,Berglund:2005dm}.

Moreover the proposal to first integrate out the heavy fields 
before adding the non-pertur\-ba\-ti\-ve potentials to the superpotential seems unnatural. 
Indeed, if one naively integrates in the heavy fields, 
inconsistencies can arise \cite{Choi:2004sx,Lust:2005dy,deAlwis:2005tf}, 
because tachyonic directions may emerge in models without 
complex structure moduli, which will be a problem after uplift to dS space.
Specifically, the moduli stabilization procedure to 
AdS vacua was studied in \cite{Denef:2005mm}
for the $T^6/Z_2\times Z_2$ orientifold, with the result that all moduli
indeed can be fixed. Moreover all other $Z_N$ and $Z_N\times Z_M$
orientfolds were studied in great detail, both at the orbifold
point \cite{Lust:2005dy} and also for blowing up the
orbifold singularities \cite{Reffert:2005mn,new}

Finally, the process of uplifting is still 
poorly understood. The uplifting by 
$\overline{D3}$-branes breaks explicitly supersymmetry, hence making a controlled uplift difficult. 
An alternative proposal is to consider D-terms due to non-supersymmetric 
2-form flux on the world-volume of D7-branes as uplifting terms \cite{Burgess:2003ic}. 
Recently, progress has been made in this direction \cite{Villadoro:2005yq,Achucarro:2006zf,Parameswaran:2006jh}.
However, one should keep in mind that all results obtained 
so far are only valid in the large volume limit such that the backreaction on the 
geometry due to the fluxes is negligible and perturbative $\alpha'$-corrections are under control.

The main focus of this work will be on 
applying the refined KKLT scenario of \cite{Blanco-Pillado:2005fn},
namely to stabilize all moduli in a Minkowski vacuum instead of
an AdS vacuum, 
to the 
orbifold models of \cite{Lust:2005dy}. 
This is interesting since several problems related to the uplift in the original
KKLT scenario can be avoided in 
%
the scheme of \cite{Blanco-Pillado:2005fn}.
In particular, whereas the
toroidal orientifold models without complex structure generally suffer from tachyonic directions in the 
minimized scalar potential after the uplift, Minkowski vacua guarantee the absence of tachyonic directions without any further input. 
As we will see   however, models without complex structure modulus
still have a problem since the axion-dilaton stays unstabilized. 
It will be found that the difficulties can be in principle solved by taking an additional effect, 
namely gaugino condensation on a stack of space-time filling fixed D3-branes into account.
Further, supersymmetric Minkowski vacua show the nice property of being qualitatively independent of perturbative corrections to the K\"ahler potential.\\

\noindent
The outline of the paper is as follows:\\
In section 2, the general conditions for a supersymmetric Minkowski vacuum are given. 
Two possible ways to fulfill the consistency condition of vanishing superpotential are discussed 
and a comment about the independence of supersymmetric Minkowski vacua on perturbative corrections to the K\"ahler potential is made.\\
Section 3 deals with microscopic details to obtain a racetrack scheme. 
It is argued that in type IIB only gaugino condensation should be a source for a racetrack potential. 
The arising difficulties in constructing a microscopic model 
are explained, and the scheme of \cite{Blanco-Pillado:2005fn} is generalized to include supersymmetric 2-form flux on D7-branes.\\
In section 4 toroidal orientifold models with one or two complex structure moduli 
in the orbifold limit are considered and it is shown that they indeed possess supersymmetric Minkowski vacua.\\
In section 5 an additional gaugino condensate on a stack of space-time 
filling D3-branes is used to construct consistent supersymmetric Minkowski vacua for orientifold models without complex structure moduli.\\
Finally, section 6 gives the conclusion.

\section{Minkowski vacua conditions}
Minkowski vacua are characterized by vanishing cosmological constant. 
For supersymmetric vacua in $\mathcal N=1$ supergravity, a vanishing cosmological constant is equivalent to a vanishing  scalar potential.\\
We limit our discussion on the F-term scalar potential, which is given by:
\begin{equation}\label{RACEeq0}
V_F=e^{K}(G^{I\bar J}D_IW\bar D_{\bar J}\bar W-3|W|^2),
\end{equation}
where $I,\bar J$ run over all moduli fields $\phi_I$, 
$K$ denotes the K\"ahler-, $W$ the superpotential and $G^{I\bar J}$ the inverse K\"ahler metric. For simplicity of notation, the set of complex structure moduli $(Z_1,...,Z_m)$ will be denoted as $Z$ and the K\"ahler moduli $(T_1,...,T_n)$ as $T$. $S$ denotes the axion-dilaton. The respective vacuum expectation values will be denoted as $T^0,S^0$ and $Z^0$.\\
The local supersymmetry conditions are given by
\begin{equation}\label{fsusy}
D_IW=\partial_IW+(\partial_IK)W=0,
\end{equation}
for all moduli $I$. At supersymmetric points, the scalar potential (\ref{RACEeq0}) reduces to:
\begin{equation}
V_F^{susy}=-3e^{K}|W(T^0,S^0,Z^0)|^2.
\end{equation}
A vanishing cosmological constant then requires
\begin{equation}\label{RACEeq1}
W(T^0,S^0,Z^0)=0. 
\end{equation}
At such points, the local supersymmetry conditions reduce to the global ones:
\begin{equation}\label{RACEeq2}
\partial_IW=0.
\end{equation}
Hence, moduli expectation values for supersymmetric Minkowski vacua can be obtained by solving (\ref{RACEeq1}) and (\ref{RACEeq2}). Eq. (\ref{RACEeq2}) can be solved in two ways: first the superpotential $W$ does not at all depend on a particular scalar field $\phi_I$, i.e. $\partial_IW\equiv 0$; this is of course not what we want,
since $\phi_I$ stays to be a flat direction in the potential. Therefore
we are looking for non-trivial solutions of eq.(\ref{RACEeq2}) with all scalar fields $\phi_I$ fixed
to specific values. As we will see this requirement may cause problems to some concrete models. Note that in contrast to the Minkowski vacua, it is not possible to get from
a F-term scalar potential non-trivial supersymmetric AdS
vacua with negative cosmological constant, which nevertheless possess still some complex flat, undetermined
moduli directions. The proof goes as follows:\\
Let $X$ be a set of moduli and $\phi$ a modulus which is a flat direction of the scalar potential, i.e. $\partial_\phi V\equiv 0$.  Further, assume that $V$ possesses an extremal point $X^0$ which stabilize the moduli $X$. If in addition the $X$ and $\phi$ satisfy the supersymmetry conditions
\begin{equation}
D_I W=\partial_IW+(\partial_IK)W=0,
\end{equation}
where $I=(X,\phi)$ at $X^0$ for all $\phi$, the flat direction of $V$ is called a supersymmetric flat direction. Note that due to this definition  the $X^0$ are necessarily independent of $\phi$.\\
If $(\partial_\phi K)|_{X^0}\neq 0$ for all $\phi$, $D_\phi W|_{X^0}=0$ requires that $W|_{X^0}\equiv(\partial_\phi W)|_{X^0}\equiv 0$ since $W$ is holomorphic and $K$ not. Hence such points are automatically Minkowski. For some $\phi$, $(\partial_\phi K)|_{(X^0,\phi)}=0$ might occur, but still $W$ needs to vanish in such points since otherwise $\phi$ would not be a flat supersymmetric direction.\\
Hence, flat complex supersymmetric directions in the scalar potential lead automatically to Minkowski vacua. Therefore, a supersymmetric AdS vacuum does not possess such flat directions and the associated unstabilized moduli.
\\\\
One immediately sees that the original KKLT scheme can not lead to supersymmetric Minkowski vacua, since the superpotential is given by
\begin{equation}
W=W_0+Ce^{-aT},
\end{equation}
where $W_0,C,a$ are constants and the second term is of 
non-perturbative origin. 
Here $T$ denotes a single K\"ahler modulus. Hence $\partial_TW=0$ can not be satisfied non-trivially for finite values of $T$. 
This changes, if one introduces additional non-perturbative $T$ dependend terms. The simplest case is the racetrack scheme
\begin{equation}
W=W_0+Ce^{-aT}-De^{-bT},
\end{equation}
with $C,D,a,b$ real positive constants. Such racetrack superpotentials with vanishing $W_0$ have already been introduced some time ago in the context of heterotic strings \cite{Krasnikov:1987jj,Dixon:1990ds,Dine:1999dx} to stabilize the dilaton and breaking supersymmetry. Lately, such potentials with non-vanishing $W_0$ gained again attention in the IIB KKLT setup \cite{Escoda:2003fa,Kallosh:2004yh,Blanco-Pillado:2004ns,Blanco-Pillado:2005fn} since they possess nice cosmological properties and a positive-definite mass matrix $\mathcal M_{I\bar J}=\partial_I\partial_{\bar J}V$ in supersymmetric Minkowski vacua, avoiding stability problems after uplifting to dS vacua. The positive-definiteness of $\mathcal M$ in supersymmetric Minkowski vacua can easily be verified, since only terms which do not involve $W$ or a first derivative of $W$ contribute to $\mathcal M$ at such vacua due to the conditions (\ref{RACEeq1}) and (\ref{RACEeq2}). Thus,
\begin{align}
\mathcal M_{MN}&=0,\label{RACEeq12}\\
\mathcal M_{\bar MN}&=e^KG^{I\bar J}(\partial_N\partial_I W)(\partial_{\bar M}\partial_{\bar J} \bar W).\label{RACEeq9}
\end{align}
Since the K\"ahler metric $G$ is positive-definite, so is $\mathcal M$.
Hence in supersymmetric Minkowski vacua, extrema of the scalar potential are always minima.\footnote{Strictly, $\mathcal M$ is only positive semi-definite, however the semi-definite case corresponds to a flat direction in the scalar potential \cite{Blanco-Pillado:2005fn}. Further note that stability of non-supersymmetric Minkowski vacua is model dependent \cite{Gomez-Reino:2006dk}.}\\\\
Another interesting possibility to obtain supersymmetric Minkowski vacua would be non-perturbative superpotentials of the following form:
\begin{equation}
W=W_0+C~T~e^{-aT},
\end{equation}
where the prefactor of the non-perturbative potential is linear in $T$.\\
For a simple one modulus system with constant flux 
superpotential $W_0$, the conditions (\ref{RACEeq1}) and (\ref{RACEeq2}) give:
\begin{equation}
T^0=\frac{1}{a},
\end{equation}
\begin{equation}
W_0=-\frac{C}{ae}.
\end{equation}
However, it is unclear if it is possible to obtain such superpotentials in a Type IIB setup, e.g.
by considering gauge threshold correction in orientifold models 
\cite{Lust:2003ky}.
It would be interesting to investigate this in future work. 
\\\\
From now on we will stick to the classical racetrack scheme, 
and assume that the geometry of the compactification manifold allows a non-perturbative potential of racetrack form for each K\"ahler modulus $T_i$:
\begin{equation}
W_{np}=\sum_i\left(C_ie^{-a_iT_i}-D_ie^{-b_i T_i}\right).
\end{equation}
Some microscopic details about such racetrack potentials in 
IIB string compactifications will be discussed in section 3. At the moment, it is just assumed that $C_i,D_i$ are positive real constants.\\
The full superpotential is then given by
\begin{equation}
W=W_{flux}+\sum_i^n \left(C_ie^{-a_iT_i}-D_ie^{-b_i T_i}\right),
\end{equation}
where $W_{flux}$ denotes the Gukov-Vafa-Witten 
superpotential arising in flux 
compactifications 
\cite{Gukov:1999ya,Taylor:1999ii,Mayr:2000hh,Giddings:2001yu}:
\begin{equation}
W_{flux}=\int_{X_6}G_{(3)}\wedge \Omega,
\end{equation}
$X_6$ denotes the compact Calabi-Yau space, 
$G_{(3)}$ is the combined 3-form flux and $\Omega$ denotes the unique globally defined harmonic (3,0)-form on $X_6$.\\
The flux potential can be parameterized as:
\begin{equation}
W_{flux}=A(Z_1,...,Z_m)+B(Z_1,...,Z_m)S,
\end{equation}
where $A,B$ are flux dependent functions.\\
The Minkowski vacuum conditions (\ref{RACEeq1}) and (\ref{RACEeq2}) lead to the following set of equations to be solved for the vacuum expectation values of the moduli:
\begin{equation}\label{RACEeq3}
T^0_i=\frac{1}{a_i-b_i}\ln\left[\frac{a_iC_i}{b_iD_i}\right],
\end{equation}
\begin{equation}\label{RACEeq4}
B(Z^0)=0,
\end{equation}
\begin{equation}\label{RACEeq5}
\partial_{Z_j}A(Z)|_{Z^0}+S^0 \partial_{Z_j}B(Z)|_{Z^0}=0,
\end{equation}
\begin{equation}\label{RACEeq6}
A(Z^0)+\omega^0=0,
\end{equation}
where $\omega^0$ has been defined as
\begin{equation}\label{RACEeq8}
\omega^0=\sum^n_i\left(C_ie^{-a_iT^0_i}-D_ie^{-b_iT^0_i}\right).
\end{equation}
This set of equations are identical to the original ones of 
\cite{Blanco-Pillado:2005fn}. 
The authors of \cite{Blanco-Pillado:2005fn} proposed to use 
equations (\ref{RACEeq4}) and (\ref{RACEeq5}) to fix the 
complex structure moduli and the axion-dilaton and to ensure by specific choice of $C_i,D_i,a_i,b_i$ that equation (\ref{RACEeq6}) is satisfied.
Alternatively to the approach by \cite{Blanco-Pillado:2005fn}, one might think about satisfying (\ref{RACEeq6}) by appropriate choice of flux.\\
Some comments are in order. Treating $C_i,D_i,a_i,b_i$ as free parameters is not necessarily justified, since $a_i,b_i,C_i,D_i$ are fixed 
by the specific compactification construction and low-energy physics 
as long as threshold corrections are neglected.\footnote{It might be 
reasonable to expect that threshold corrections \cite{Lust:2003ky} will lead at least 
to a complex structure dependence of the gauge kinetic function which can be seen as a complex structure dependence of the prefactors $C_i,D_i$.}
However, solving (\ref{RACEeq6}) by tuning of fluxes is also not necessarily possible, since flux can only be tuned discreetly. Nevertheless, since it is the simplest approach, in the following we will assume that flux degrees of freedom can be chosen such that (\ref{RACEeq6}) is satisfied, keeping in mind that this may not always be possible. \\
Also note that the vacuum expectation values of the moduli are stable against perturbative corrections to the K\"ahler potential, since the expectation values are completely determined by the superpotential. The same holds for the positive-definiteness property of the mass matrix since (\ref{RACEeq9}) only depends on the K\"ahler potential via the $e^K$ prefactor and the K\"ahler metric which stays positive definite under perturbative corrections.\\\\

\section{Racetrack potentials}
\begin{figure}
\begin{center}
\psfrag{T}[br]{\small $T$}
\psfrag{X}[bl]{\small $\Theta$}
\psfrag{N}[bl]{\small $N$}
\includegraphics[scale=0.8]{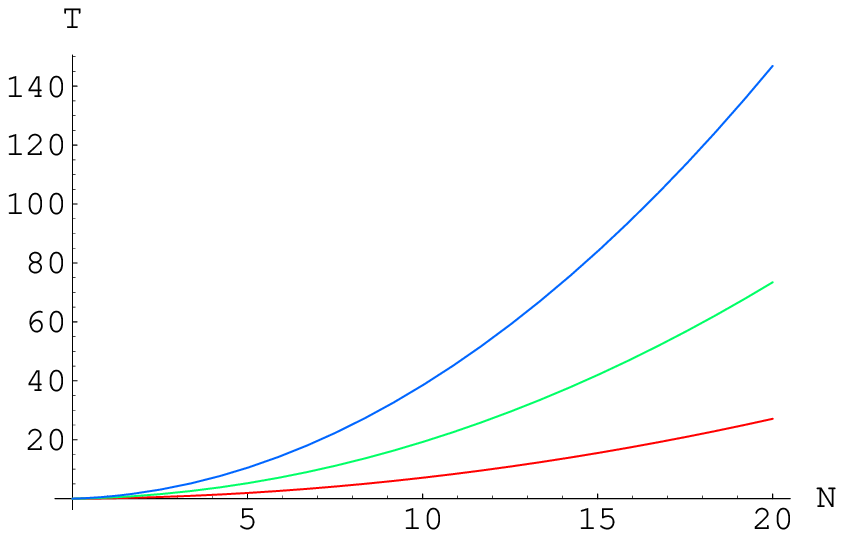}
\hspace{1cm}
\includegraphics[scale=0.8]{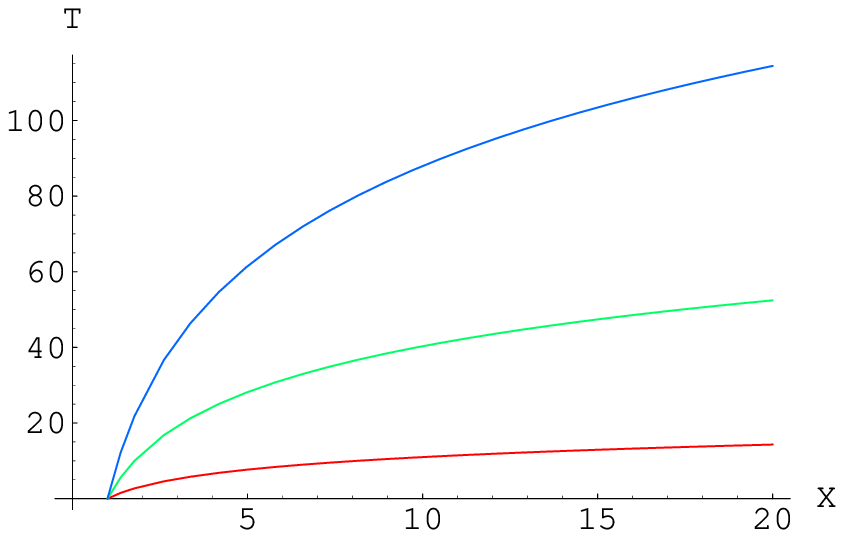}
\caption[Tmoduli]{Left: T moduli in dependence of $N_c$ for $\Theta=1.2$ (red line), $\Theta=3$ (green line), $\Theta=9$ (blue line). Right: T moduli in dependence of $\Theta$ for $N_c=5$ (red line), $N_c=10$ (green line), $N_c=15$ (blue line).}
\label{RACEfig1}
\end{center}
\end{figure}

There are two known sources of possible non-perturbative corrections to superpotentials in Type IIB orientifold compactifications. Namely, instanton effects due to euclidean D3-branes wrapping four cycles in $X_6$ and gaugino condensation in supersymmetric gauge theories on the worldvolume of D-branes.\\
The instantons yield the following non-perturbative superpotential \cite{Witten:1996bn}:
\begin{equation}
W_{np}=C(Z)e^{-2\pi T},
\end{equation}
where $C(Z)$ is a complex structure dependent one-loop determinant and $T$ the K\"ahler modulus associated to the volume of the 4-cycle wrapped by the euclidean D3-branes. The explicit form of $C(Z)$ is unknown in general. Generally, the existence of such instantons is only possible if the 4-cycle satisfies certain topological properties. \\
It is reasonable to expect that a racetrack potential can not be generated by two instantons on the same cycle, since in this case the two non-perturbative terms should combine to a single KKLT type term. 
Also D3 branes and higher dimensional D-branes on the same cycle should combine to a single stack of D-branes, preventing the simultaneous creation of an instanton and a gaugino condensate.\\
For these reasons, only gaugino condensation might be seen as a candidate for generating racetrack potentials.\\
Gaugino condensation is a low energy effect in supersymmetric gauge theories. If no additional matter is present (pure super Yang-Mills), the following non-perturbative superpotential is generated: 
\begin{equation}\label{RACEeq10}
W_{np}\sim b e^{-\frac{3}{2b} f},
\end{equation}
where $b$ is the $\beta$-function coefficient of the gauge group and $f$ the gauge kinetic function. For gauge theories on the world-volume of D-branes, the gauge kinetic functions are related to moduli. Of special interest are stacks of D3 and D7-branes, since these can occur simultaneously in a supersymmetric IIB orientifold compactification. To first order, one finds for gauge theories on stacks of D7 branes filling space-time and wrapping a 4-cycle of $X_6$:
\begin{equation}
f_{D7}=T,
\end{equation}
while for gauge theories on stacks of space-time filling D3 branes:
\begin{equation}
f_{D3}=S.
\end{equation}
For pure $SU(N)$ super Yang-Mills, $b$ is given by the quadratic Casimir of $SU(N)$. In this case, the non-perturbative potential is given by:
\begin{equation}
W_{np}=NCe^{-\frac{2\pi}{N}f},
\end{equation}
where $C$ is an $O(1)$ constant determined by low-energy physics and $N$ is the rank of $SU(N)$.\\
The existence of such gaugino condensates giving non-perturbative potentials for the 
K\"ahler moduli of the orientifold, puts strong constraints on the topology of $X_6$. 
The constraints arising for toroidal orientifolds were discussed in \cite{Lust:2005dy}. Similar constraints results hold in more general orientifolds. In detail, additional fundamental, bi-fundamental and adjoint matter, due to intersecting D-Branes, Wilson-lines and/or variable D-brane positions, may spoil the gaugino condensate. One must make sure that such additional matter does not exist or becomes massive, e.g. by appropriately switching on 3-form fluxes.\\
In order to obtain a racetrack scheme, one needs to break the gauge group $\mathcal G$ of a stack of $N$ D7-branes wrapping a 4-cycle down to a product gauge group $\mathcal{G}_c\times \mathcal{G}_d$, such that gaugino condensation occurs independently in each gauge sector. The standard procedure to achieve this, is to use the translational degrees of freedom in the transversal space of the 4-cycle to fix $N_c$ and $N_d$ branes (with $N_c+N_d=N$) at different positions in the transversal space. If the 4-cycle of $X_6$ has no transversal degrees of freedom, one needs to break the gauge group by Wilson-lines or by switching on different 2-form flux on the branes. Note that for the first two possibilities, the structure of the broken gauge group is determined by the vacuum expectation values of the associated scalar fields associated to the positions of the D7-branes and Wilson-lines. Hence, 3-form flux must be properly switched on such that these fields have no flat directions in the scalar potential such that they become massive so that gaugino condensation can occur.\\
If one switches on in addition 2-form fluxes on the worldvolume of the D7-branes in toroidal orientifolds, 
the gauge kinetic function $f_{D7}$ changes as 
follows \cite{Lust:2004cx,Lust:2004fi}:\footnote{Racetrack models with similar gauge coupling, but with $W_{flux}=0$ were considered in \cite{Abe:2005pi,Abe:2005rx}.} 
\begin{equation}
f_{D7}=T-\gamma S,
\end{equation}
where $\gamma$ is a complex constant parameterizing the 2-form flux. 
One should keep in mind that due to the switched on 2-form flux, also D7-branes can contribute to the Ramond-Ramond 4-form tadpole conditions. 
In addition it is assumed that the 2-form flux preserves supersymmetry, which means that the associated D-term potential is vanishing. \\
There is also the possibility that additional moduli dependence may occur due to threshold corrections to the gauge kinetic function. However, stabilizing the axion-dilaton and volume sufficiently large, the freedom might be taken to neglect them as is done in the following.\\\\
If the $X_6$ under consideration supports a consistent D7-brane setup which leads to such a pure super Yang-Mills with product gauge group $SU(N_c)\times SU(N_d)$ for each K\"ahler modulus, gaugino condensation in both gauge sectors will give the following superpotential if no 2-form flux is switched on:
\begin{equation}
W=W_{flux}+\sum_i^n\left(N_c^iC_ie^{-\frac{2\pi}{N_c^i}T_i}-N_d^iD_ie^{-\frac{2\pi}{N_d^i}T_i}\right).
\end{equation}
In this setup, equation (\ref{RACEeq3}) reads
\begin{equation}
T^0_i=\frac{1}{2\pi}\frac{N_c^iN_d^i}{N_d^i-N_c^i}\ln\left[\frac{C_i}{D_i}\right].
\end{equation}
Note that for real $C_i,D_i$ it is necessary that the prefactors of both non-perturbative terms differ in sign. Since the gauge interactions do not fix the phase of gaugino condensates \cite{Krasnikov:1987jj,Casas:1990qi}, this should be possible as long as the rank of one of the gauge groups is even.
One immediately sees that $N^i_c \neq N_d^i$ is required. Otherwise one is back to the standard KKLT scheme. Also, the positive-definiteness of the K\"ahler moduli require one of the two following conditions to be satisfied:
\begin{align}
N^i_d>N^i_c,\ C^i>D^i
\end{align}
or
\begin{align}
N^i_d<N^i_c,\ C^i<D^i.
\end{align}
In the following it will be assumed that the parameters satisfy the first case.\\
One immediately sees that for realistic gauge group ranks a stabilization at large $T_i$ values requires $N^i_c$ to be close to $N^i_d$.\\
The largest possible $T_i$ value can be obtained for $N^i_d=N^i_c+1$:
\begin{equation}\label{RACEeq7}
T^0_i=\frac{1}{2\pi}N_c^i(N_c^i+1)\ln\left[\Theta_i\right],
\end{equation}
with 
\begin{equation}
\Theta_i=\frac{C_i}{D_i}.
\end{equation}
$T_i$ in dependence of $N^i_c$ for several values of $\Theta_i$ and $T_i$ in dependence of $\Theta_i$ for several values of $N^i_c$ is plotted in figure \ref{RACEfig1}.\\ 
Clearly, a stabilization at large volume requires that $N^i_c$ and $\Theta_i$ are large. This may become problematic for resolved toroidal orientifold models, since these generally possess a large amount of K\"ahler moduli. As argued before, only gaugino condensation should be a source for racetrack potentials and hence every 4-cycle must be wrapped with a stack of approximately twenty D7-Branes to achieve a large volume supersymmetric Minkowski vacua. In total, it is reasonable to expect that several hundreds of D7 branes are needed to obtain such vacua via Racetrack potentials, making the cancellation of D7 charge and also Ramond-Ramond 4-form charge tadpole conditions difficult.\\
For later convenience, define as in (\ref{RACEeq8}) $\omega^0=W_{np}(T^0)$:
\begin{equation}
\omega^0=\sum_iD_i\Theta_i^{-N_c^i}.
\end{equation}
\\In the identical setup, but with 2-form flux turned on, the superpotential becomes:
\begin{equation}
W=W_{flux}+\sum_i^n\left(N_c^iC_ie^{-\frac{2\pi}{N_c^i}T_i+\Lambda^i_c}-N_d^iD_ie^{-\frac{2\pi}{N_d^i}T_i+\Lambda^i_d}\right),
\end{equation}
with
\begin{equation}
\Lambda^i_l=\frac{2\pi}{N^i_l}\gamma^iS,
\end{equation}
for $l=c,d$. Note that necessarily $\Lambda^i_c\neq \Lambda^i_d$ since $N_c\neq N_d$ and that it was assumed that the flux factor $\gamma$ is identical for both gauge sectors. Hence it is assumed that the gauge group is broken by translation in transversal space or by Wilson-lines.\\
The K\"ahler moduli are stabilized at:
\begin{equation}
\begin{split}
T^0_i&=\frac{1}{2\pi}N_c^i(N_c^i+1)\left(\ln\left[\Theta_i\right]+(\Lambda^i_c-\Lambda^i_d)\right)\\
&=\frac{1}{2\pi}N_c^i(N_c^i+1)\ln\left[\Theta_i\right]+\gamma^iS^0.
\end{split}
\end{equation}
Observe that the vacuum expectation value of the non-perturbative superpotential with 2-form flux $W_{np}(T^0,S^0)$ equals the vacuum expectation value of the non-perturbative superpotential without 2-form flux $\omega^0$. Further, $(\partial_SW_{np}(T,S))|_{T^0}$ vanishes.\footnote{Note that this only holds for $N_d=N_c+1$. Therefore we will always consider this case in the following models if two-form flux is switched on.} Hence, the set of equations (\ref{RACEeq4})-(\ref{RACEeq6}) is identical for models with and without supersymmetric 2-form flux. The only difference is that the vacuum expectation values of the K\"ahler moduli get an additional flux and axion-dilaton dependent term.\\
Thus, the generalized set of equations determining the vacuum expectation values of the moduli with and without 2-form flux in a Minkowski vacuum for $SU(N^i_c)\times SU(N^i_c+1)$ gauge theories living on D7-branes is given by:
\begin{equation}\label{3RACEeq1}
\begin{split}
&T^0_i=\frac{1}{2\pi}N_c^i(N_c^i+1)\ln\left[\Theta_i\right]+\gamma^iS^0,\\
&B(Z^0)=0,\\
&(\partial_{Z_j}A(Z))|_{Z^0}+S^0(\partial_{Z_j}B(Z))|_{Z^0}=0,\\
&A(Z^0)+\omega^0=0.
\end{split}
\end{equation}
The D7 charge cancellation in this setup is less problematic, since 
fewer D7-branes are needed for a large volume stabilization if 2-form flux 
is properly switched on. In sections 4 and 5 this scheme will be explicitly applied to some toroidal orientifold models.

\section{Toroidal orientifold models with complex structure moduli (CSM)}
\subsection{One CSM}
In \cite{Lust:2005dy} AdS moduli stabilization in $Z_N$ and $Z_N\times Z_M$ orientifold models
was discussed. Here were are interested in the question, whether these
models can also lead to Minkowski vacua.
If the consistency conditions discussed in section 3 are satisfied, 
the models with $h_{(1,1)}=n$, $h^{untw}_{(2,1)}=1$, 3-form flux and additional 2-form flux possess the following superpotential:
\begin{equation}
W=(\alpha_1+\alpha_2Z)+(\alpha_3+\alpha_4Z)S+\sum_i^n\left(N_c^iC_ie^{-\frac{2\pi}{N^i_c}T_i+\Lambda^i_c}-N_d^iD_ie^{-\frac{2\pi}{N^i_d}T_i+\Lambda^i_d}\right).
\end{equation}
This captures the $\mathbb Z_{6-II},\mathbb Z_2\times \mathbb Z_3,\mathbb Z_2\times \mathbb Z_6$ models in the orbifold limit and also the $\mathbb Z_{6-II'}$ model after blowup since $h^{twist}_{2,1}=0$ \cite{Lust:2005dy}.\\
For convenience, the 3-form flux matrix $G_3$ will be defined as
\begin{equation}
G_3=
\begin{pmatrix}
\alpha_1 & \alpha_2\\
\alpha_3 & \alpha_4\\
\end{pmatrix},
\end{equation}
where $\alpha_i$ are 3-form flux dependent constants.\\
Application of equations (\ref{3RACEeq1}) gives:
\begin{equation}
\begin{split}
Z^0&=-\frac{\alpha_3}{\alpha_4},\\
S^0&=-\frac{\alpha_2}{\alpha_4},\\
\alpha_1&=-\alpha_2Z^0-\omega^0.
\end{split}
\end{equation}
Substitution of the first equation into the last gives the condition:
\begin{equation}
\det(G_3)=-\alpha_4\omega^0.
\end{equation}
A choice of 3-form flux
\begin{equation}
G_3=
\begin{pmatrix}
-\omega^0-\alpha_2\alpha_3 & \alpha_2,\\
\alpha_3 & -1\\
\end{pmatrix},
\end{equation}
gives a consistent supersymmetric Minkowski vacuum with tuneable $S^0$ and $Z^0$:
\begin{equation}
\begin{split}
S^0&=\alpha_2\\
Z^0&=\alpha_3.
\end{split}
\end{equation}
The fixed K\"ahler moduli are given in (\ref{RACEeq7}).

\subsection{Two CSM}
Toroidal orientifolds with $h_{(1,1)}=n$ and $h^{untw}_{(2,1)}=2$ which fulfill tadpole conditions do not exist. However, if one identifies two of the complex structure moduli this case captures the $\mathbb Z_2\times \mathbb Z_2$ model. For simplicity, we will stick to this case. Since the $\mathbb Z_2\times \mathbb Z_2$ model has $h^{twist}_{(2,1)}=0$, the resolved case \cite{Denef:2005mm} is captured as well.\\  
The superpotential is given by:
\begin{equation}
\begin{split}
W=&(\alpha_1+\alpha_2Z_1)+(\alpha_3+\alpha_4Z_1)S+(\alpha_5+\alpha_6S)Z_2+(\alpha_7+\alpha_8S)Z_1Z_2\\
&+\sum_i^n\left(N^i_cC_ie^{-\frac{2\pi}{N^i_c}T_i+\Lambda^i_c}-N_d^iD_ie^{-\frac{2\pi}{N^i_d}T_i+\Lambda^i_d}\right).
\end{split}
\end{equation}
Again a racetrack type superpotential with possible 2-form flux was taken into account.\\
For convenience, the 3-form flux matrix $G_3$ will be defined as
\begin{equation}
G_3=
\begin{pmatrix}
\alpha_1&\alpha_2&\alpha_3\\
\alpha_4&\alpha_5&\alpha_6\\
\alpha_7&\alpha_8&0\\
\end{pmatrix}.
\end{equation}
Equations (\ref{3RACEeq1}) lead to:
\begin{equation}
\begin{split}
\alpha_3+\alpha_4Z^0_1+\alpha_6Z^0_2+\alpha_8Z^0_1Z^0_2&=0,\\
\alpha_2+\alpha_4S^0+\alpha_7Z^0_2+\alpha_8S^0Z^0_2&=0,\\
\alpha_5+\alpha_6S^0+\alpha_7Z^0_1+\alpha_8S^0Z^0_1&=0,\\
\alpha_1+\alpha_2Z^0_1+\alpha_5Z^0_2+\alpha_7Z^0_1Z^0_2+\omega^0&=0.
\end{split}
\end{equation}
The first three equations simplify to:
\begin{equation}
S^0=-\frac{\alpha_5+\alpha_7Z^0_1}{\alpha_6+\alpha_8Z^0_1},\ Z^0_1=-\frac{\alpha_3+\alpha_6Z^0_2}{\alpha_4+\alpha_8Z^0_2},\ Z^0_2=-\frac{\alpha_2+\alpha_4S^0}{\alpha_7+\alpha_8S^0}.
\end{equation}
If the flux parameters satisfy certain determinant conditions, the vacuum expectation values of the moduli are related among each other by projective conformal transformations given by the group $PSL(2,\mathbb C)$.\\
Choosing the 3-form flux as
\begin{equation}
G_3=
\begin{pmatrix}
-(\omega^0+\frac{\alpha_5}{2})&0&0\\
1&\alpha_5&1\\
1&0&0\\
\end{pmatrix},
\end{equation}
gives:
\begin{align}
Z^0_1&=S^0,\\
Z^0_2&=-S^0,\\
S^0&=-\frac{\alpha_5}{2}.
\end{align}
Hence a consistent supersymmetric Minkowski vacuum with tuneable $S^0$. The fixed K\"ahler moduli are given in (\ref{RACEeq7}).

\section{Models without CSM}
The general racetrack superpotential for orientifolds with $h_{(1,1)}=n$, $h_{(2,1)}=0$ and possible 2-form flux is given by:
\begin{equation}
W=\alpha_1+\alpha_2S+\sum^n_i\left(N^i_cC_ie^{-\frac{2\pi}{N_c}T_i+\Lambda^i_c}-N^i_dD_ie^{-\frac{2\pi}{N_d}T_i+\Lambda^i_d}\right),
\end{equation}
where $\alpha_i$ are complex constants determined by 3-form fluxes.\\
Since it is reasonable to expect that $W$ holds even after 
blowing up toroidal orbifolds, as long as $h^{twist}_{2,1}=0$, 
this specially includes the $\mathbb Z_3,\mathbb  Z_7,\mathbb  Z_3\times\mathbb  Z_3,\mathbb  Z_6\times \mathbb Z_6$ and $\mathbb Z_2\times\mathbb  Z_{6'}$ toroidal orientifold models before and after blowup and the $\mathbb Z_{6-I},\mathbb Z_{12-I}$ and $\mathbb Z_3\times\mathbb  Z_6$ models in the orbifold limit \cite{Lust:2005dy}.\\ 
The Minkowski vacua condition $\partial_SW=0$ immediately shows that one 
is forced to set $\alpha_2=0$. 
In this case, $S$ is a flat direction of the superpotential.\footnote{Strictly, 
this is only valid in the case when the 2-form flux $\gamma$ is identical for both gauge sectors, 
since only in this case the additional axion-dilaton dependence due to the 2-form fluxes vanishes in the vacuum.} The axion-dilaton stays unstabilized.\\ 
Hence, the scheme fails for models without complex structure moduli.
However, if an additional stack of fixed D3 branes is present, gaugino condensation can occur in the corresponding gauge theory with gauge coupling 
\begin{equation}
f_{D3}=S.
\end{equation}
This leads to an additional term to the non-perturbative superpotential of the form
\begin{equation}
W_{D3}=N_eEe^{-\frac{2\pi}{N_e}S},
\end{equation}
where $E$ is an $O(1)$ constant.\\
The condition $\partial_SW=0$ then gives
\begin{equation}
S^0=\frac{N_e}{2\pi}\ln\left[\frac{2\pi E}{\alpha_2}\right].
\end{equation}
Positive definiteness of $S^0$ requires that
\begin{equation}
\alpha_2<2\pi E.
\end{equation}
The consistency condition $W(T^0,S^0)=0$ can be fulfilled by setting
\begin{equation}
\alpha_1=-\left(\left(S^0+\frac{N_e}{2\pi}\right)\alpha_2+\omega^0\right),
\end{equation}
The $T^0_i$ values are unaffected and given in (\ref{RACEeq7}).\\
Hence, with this modified scheme it is possible to stabilize to a supersymmetric Minkowski vacuum with tuneable $S^0$. \\
For illustration, the scalar potential (\ref{RACEeq0}) is plotted in figure \ref{RACEfig2} for a sample choice of purely real parameters using the standard K\"ahler potential with identified K\"ahler moduli
\begin{equation}\label{RACEeq11}
K=-n\ln(T+\bar T)-\ln(S+\bar S),
\end{equation}
valid for the mentioned $\mathbb Z_7,\mathbb Z_{12-I},\mathbb Z_3\times\mathbb Z_3,\mathbb Z_6\times\mathbb Z_6,\mathbb Z_3\times\mathbb Z_6$ and $\mathbb Z_3\times\mathbb Z_{6'}$ models in orbifold limit.
\begin{figure}
\label{RACEfig2}
\begin{center}
\psfrag{V}[br]{\small $V$}
\psfrag{s}[br]{\small $s$}
\psfrag{t}[br]{\small $t$}
\includegraphics[scale=0.8]{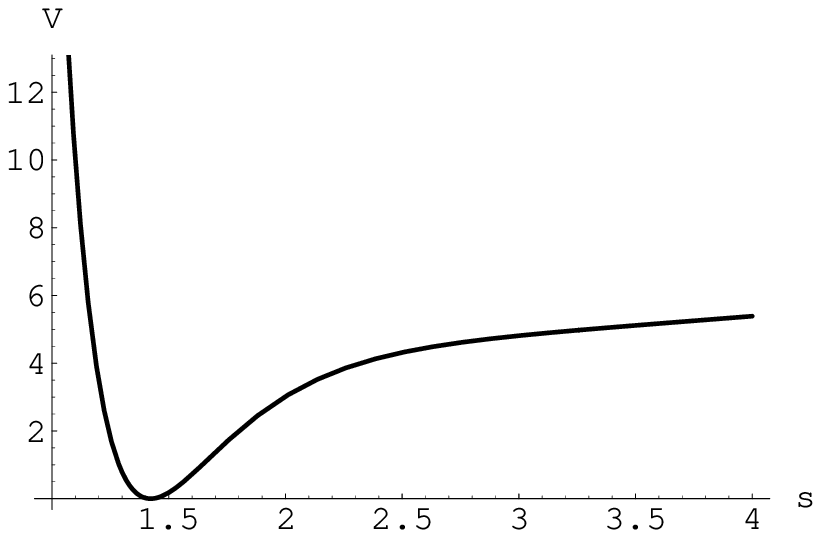}
\hspace{1cm}
\includegraphics[scale=0.8]{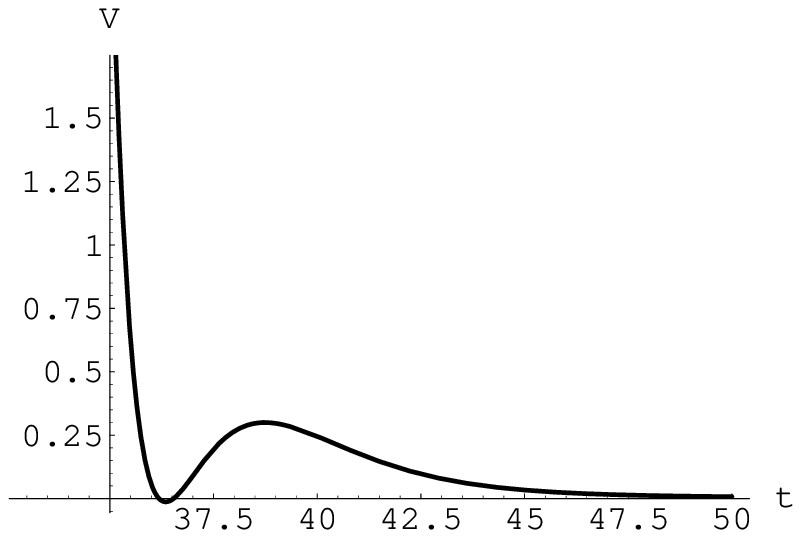}
\caption[Race3K0CSM]{Slides of the F-Term scalar potential for models without complex structure, K\"ahlerpotential as in (\ref{RACEeq11}), racetrack potential (without 2-form flux) and D3 brane gaugino condensation.\\Left: V with K\"ahler moduli fixed at $t\approx T^0$ (V is multiplied by $10^{8}$),
\\Right: V with axion-dilaton fixed at $s\approx S^0$ (V is multiplied by $10^{16}$).\\ Choice of parameters as follows: $n=3$, $N_c=14$, $N_d=15$, $N_e=2$, $C=3$, $D=1$, $F=1$, $\alpha_2\approx 0.072$.}
\label{RACEfig2}
\end{center}
\end{figure}
\\Note that this scheme gives the first realization of moduli stabilization 
without tachyonic directions in toroidal orientifold models without complex structure and without first integrating out the axion-dilaton.

\section{Conclusion}
In this paper we discussed the possibility to get Minkowski vacua in
type IIB orientifold models with all moduli stabilized.
We first showed that there are two serious obstacles for explicit IIB 
orientifold Minkowski vacua. Besides the complications to be overcome for gaugino 
condensation to occur, the situation becomes even more problematic since 
Minkowski vacua require that the consistency condition $W(T^0,Z^0,S^0)=0$ is 
fulfilled. There are two possibilities to achieve this. Tuning of 3-form 
fluxes or restrictions on parameters $a_i,b_i,C_i,D_i$. Tuning of fluxes 
gives a convenient way to fulfill the consistency condition, however 
with the major drawback that this may not be possible for every model 
since flux is only tunable discretely. The second possibility, 
treating the parameters $a_i,b_i,C_i,D_i$ as free, 
means further strong restrictions on the compactification manifold, 
since $a_i,b_i$ depend on the number of D-branes and $C_i,D_i$ are determined by 
low-energy physics as long as one does not take threshold effects into account. 
Hence, no matter which way one chooses to fulfill the consistency condition, 
finding an explicit compactification setup is much more restrictive and difficult than in the standard KKLT scheme.

However in case being realized,
Minkowski vacua in IIB orientifold compactifications offer very nice phenomenological features. 
Besides the properties already observed in 
\cite{Kallosh:2004yh,Blanco-Pillado:2005fn}, namely the possibility to have 
high energy scale inflation with low-energy supersymmetry breaking and solution of stability problems in the uplifting process, 
two new features were observed. Firstly, that properly switched on 2-form flux only affects the K\"ahler moduli 
vacuum expectation values. Secondly, stability against perturbative corrections 
to the K\"ahler potential of the moduli 
vacuum expectation values and of the positive-definiteness property of the 
mass matrix.

While tachyonic directions are automatically absent in supersymmetric Minkowsi vacua, flat directions may occur. Interestingly, supersymmetric AdS vacua show the opposite properties. In a concrete model without complex structure the flat direction can be lifted by taking an additional effect into account, namely gaugino condensation on space-time filling D3-branes.
 
Finally note that also in KKLT scenarios with AdS vacua additional
gaugino condensation on fixed D3 branes may solve the stability problems \cite{Choi:2004sx,Choi:2005ge,Lust:2005dy} in models without complex structure 
modulus\footnote{A similar 
observation was mentioned in a footnote of \cite{Conlon:2005ki}}.

\subsection*{Acknowledgements}
We would like to thank 
M. Haack, S. Reffert and S. Stieberger for many valuable discussions. 
This work is partly supported by EU contract MRTN-CT-2004-005104.


\begin{thebibliography}{77}
\bibitem{Grana:2005jc}
M.~Grana,
Phys.\ Rept.\  {\bf 423}, 91 (2006)
[arXiv:hep-th/0509003].
\bibitem{Kachru:2003aw}
  S.~Kachru, R.~Kallosh, A.~Linde and S.~P.~Trivedi,
  Phys.\ Rev.\ D {\bf 68} (2003) 046005
  [arXiv:hep-th/0301240].
\bibitem{vonGersdorff:2005bf}
  G.~von Gersdorff and A.~Hebecker,
  Phys.\ Lett.\ B {\bf 624} (2005) 270
  [arXiv:hep-th/0507131].
\bibitem{Balasubramanian:2005zx}
V.~Balasubramanian, P.~Berglund, J.~P.~Conlon and F.~Quevedo,
JHEP {\bf 0503}, 007 (2005)
[arXiv:hep-th/0502058].
\bibitem{Berg:2005yu}
  M.~Berg, M.~Haack and B.~K\"ors,
  arXiv:hep-th/0508171.
\bibitem{Witten:1996bn}
  E.~Witten,
  Nucl.\ Phys.\ B {\bf 474}, 343 (1996)
  [arXiv:hep-th/9604030].
\bibitem{Gorlich:2004qm}
  L.~G\"orlich, S.~Kachru, P.~K.~Tripathy and S.~P.~Trivedi,
  JHEP {\bf 0412} (2004) 074
  [arXiv:hep-th/0407130].
\bibitem{Kallosh:2005gs}
R.~Kallosh, A.~K.~Kashani-Poor and A.~Tomasiello,
JHEP {\bf 0506}, 069 (2005)
[arXiv:hep-th/0503138].
\bibitem{Martucci:2005rb}
L.~Martucci, J.~Rosseel, D.~Van den Bleeken and A.~Van Proeyen,
Class.\ Quant.\ Grav.\  {\bf 22}, 2745 (2005)
[arXiv:hep-th/0504041].
\bibitem{Bergshoeff:2005yp}
  E.~Bergshoeff, R.~Kallosh, A.~K.~Kashani-Poor, D.~Sorokin and A.~Tomasiello,
  JHEP {\bf 0510} (2005) 102
  [arXiv:hep-th/0507069].
\bibitem{Berglund:2005dm}
  P.~Berglund and P.~Mayr,
  arXiv:hep-th/0504058.
\bibitem{Choi:2004sx}
  K.~Choi, A.~Falkowski, H.~P.~Nilles, M.~Olechowski and S.~Pokorski,
  JHEP {\bf 0411}, 076 (2004)
  [arXiv:hep-th/0411066].
\bibitem{Lust:2005dy}
  D.~L\"ust, S.~Reffert, W.~Schulgin and S.~Stieberger,
  arXiv:hep-th/0506090.
\bibitem{deAlwis:2005tf}
  S.~P.~de Alwis,
  Phys.\ Lett.\ B {\bf 626} (2005) 223
  [arXiv:hep-th/0506266].
\bibitem{Denef:2005mm}
F.~Denef, M.~R.~Douglas, B.~Florea, A.~Grassi and S.~Kachru,
arXiv:hep-th/0503124.
\bibitem{Reffert:2005mn}
S.~Reffert and E.~Scheidegger,
arXiv:hep-th/0512287.
\bibitem{new}
D.~L\"ust, S.~Reffert, E.~Scheidegger, W.~Schulgin
and S.~Stieberger, paper to appear.
\bibitem{Burgess:2003ic}
  C.~P.~Burgess, R.~Kallosh and F.~Quevedo,
  JHEP {\bf 0310}, 056 (2003)
  [arXiv:hep-th/0309187].
\bibitem{Villadoro:2005yq}
  G.~Villadoro and F.~Zwirner,
  Phys.\ Rev.\ Lett.\  {\bf 95} (2005) 231602
  [arXiv:hep-th/0508167].
\bibitem{Achucarro:2006zf}
  A.~Achucarro, B.~de Carlos, J.~A.~Casas and L.~Doplicher,
  arXiv:hep-th/0601190.
\bibitem{Parameswaran:2006jh}
  S.~L.~Parameswaran and A.~Westphal,
  arXiv:hep-th/0602253.
\bibitem{Blanco-Pillado:2005fn}
  J.~J.~Blanco-Pillado, R.~Kallosh and A.~Linde,
  arXiv:hep-th/0511042.
\bibitem{Krasnikov:1987jj}
  N.~V.~Krasnikov,
  Phys.\ Lett.\ B {\bf 193} (1987) 37.
\bibitem{Dixon:1990ds}
  L.~J.~Dixon,
SLAC-PUB-5229
{\it Invited talk given at 15th APS Div. of Particles and Fields General Mtg., Houston,TX, Jan 3-6, 1990}
\bibitem{Dine:1999dx}
  M.~Dine and Y.~Shirman,
  Phys.\ Rev.\ D {\bf 63} (2001) 046005
  [arXiv:hep-th/9906246].
\bibitem{Escoda:2003fa}
  C.~Escoda, M.~Gomez-Reino and F.~Quevedo,
  JHEP {\bf 0311}, 065 (2003)
  [arXiv:hep-th/0307160].
\bibitem{Kallosh:2004yh}
  R.~Kallosh and A.~Linde,
  JHEP {\bf 0412} (2004) 004
  [arXiv:hep-th/0411011].
\bibitem{Blanco-Pillado:2004ns}
  J.~J.~Blanco-Pillado {\it et al.},
  JHEP {\bf 0411}, 063 (2004)
  [arXiv:hep-th/0406230].
\bibitem{Gomez-Reino:2006dk}
  M.~Gomez-Reino and C.~A.~Scrucca,
  arXiv:hep-th/0602246.
\bibitem{Lust:2003ky}
D.~L\"ust and S.~Stieberger,
arXiv:hep-th/0302221.
\bibitem{Gukov:1999ya}
  S.~Gukov, C.~Vafa and E.~Witten,
  Nucl.\ Phys.\ B {\bf 584} (2000) 69
  [Erratum-ibid.\ B {\bf 608} (2001) 477]
  [arXiv:hep-th/9906070].
\bibitem{Taylor:1999ii}
T.~R.~Taylor and C.~Vafa,
Phys.\ Lett.\ B {\bf 474}, 130 (2000)
[arXiv:hep-th/9912152].
\bibitem{Mayr:2000hh}
P.~Mayr,
Nucl.\ Phys.\ B {\bf 593}, 99 (2001)
[arXiv:hep-th/0003198].
\bibitem{Giddings:2001yu}
S.~B.~Giddings, S.~Kachru and J.~Polchinski,
Phys.\ Rev.\ D {\bf 66}, 106006 (2002)
[arXiv:hep-th/0105097].
\bibitem{Lust:2004cx}
D.~L\"ust, P.~Mayr, R.~Richter and S.~Stieberger,
Nucl.\ Phys.\ B {\bf 696}, 205 (2004)
[arXiv:hep-th/0404134].
\bibitem{Lust:2004fi}
  D.~L\"ust, S.~Reffert and S.~Stieberger,
  Nucl.\ Phys.\ B {\bf 706} (2005) 3
  [arXiv:hep-th/0406092].
\bibitem{Casas:1990qi}
  J.~A.~Casas, Z.~Lalak, C.~Munoz and G.~G.~Ross,
  Nucl.\ Phys.\ B {\bf 347} (1990) 243.
\bibitem{Abe:2005pi}
  H.~Abe, T.~Higaki and T.~Kobayashi,
  arXiv:hep-th/0512232.
\bibitem{Abe:2005rx}
  H.~Abe, T.~Higaki and T.~Kobayashi,
  arXiv:hep-th/0511160.
\bibitem{Conlon:2005ki}
  J.~P.~Conlon, F.~Quevedo and K.~Suruliz,
  JHEP {\bf 0508} (2005) 007
  [arXiv:hep-th/0505076].
\bibitem{Choi:2005ge}
  K.~Choi, A.~Falkowski, H.~P.~Nilles and M.~Olechowski,
  Nucl.\ Phys.\ B {\bf 718} (2005) 113
  [arXiv:hep-th/0503216].
\end{thebibliography}
\end{document}